\documentclass[fleqn,10pt]{wlscirep}
\usepackage{amsfonts}
\usepackage{color}
\usepackage{epstopdf}

\title{The effects of three-dimensional defects on one-way surface plasmon propagation for photonic topological insulators comprised of continuous media}

\author[1,*]{S. Ali Hassani Gangaraj}
\author[1]{Andrei Nemilentsau}
\author[1]{George W. Hanson}

\affil[1]{Department of Electrical Engineering and Computer Science, University of Wisconsin-Milwaukee, Milwaukee, Wisconsin 53211, USA}

\affil[*]{Hassani4@uwm.edu}

\begin{abstract}
We have investigated the one-way surface plasmon-polariton (SPP) at the interface of a continuous magneto-plasma material and metal. We demonstrated that TM modes inside a continuous material can be assigned non-trivial Chern numbers analogous to those of topological photonic crystals; moreover these Chern numbers can be calculated analytically. This leads to the appearance of topologically protected surface modes propagating at frequencies inside the bandgap of the magneto-plasma. Previous works considered 2D structures; here we consider the effects of 3D defects, and show that, although backward propagation/reflection cannot occur, side scattering does take place and it has significant effect on the propagation of the surface mode. Several different waveguiding geometries are considered for reducing the effects of side-scattering, and we also consider the effects of metal loss. 

\end{abstract}
\begin{document}

\flushbottom
\maketitle
%
%
\thispagestyle{empty}

\section*{Introduction}

Extensive research has been carried out in the field of surface plasmon-polaritons (SPP) due to their technological potential and fundamental nature \cite{Homola,William}. Recently, exciting phenomena have been studied such as unidirectional plasmon coupling \cite{Lopez}, plasmon focusing \cite{Yin,Nomura}, waveguiding and interferometry \cite{Bozhevolnyi,G} and planar optical chirality \cite{Zheludev,Krasavin}. For example, it has been shown that SPP waves can be excited in a single direction using a circularly-polarized source that couples to SPP spin polarization \cite{Gorodetski}. Upon encountering a discontinuity, partial reflection of the wave will occur since the material itself allows propagation in both directions. However, if the medium only supports modes that can travel in one direction, then upon encountering a discontinuity an SPP cannot be reflected (back-scattered). This is a rather remarkable occurrence, and has important applications in waveguiding (e.g., defect-immune waveguides). 

More generally, a broad class of materials known as photonic topological insulators (PTIs) provide backscattering-immune unidirectional mode propagation \cite{Haldane, Haldane2, Wang, Poo, Fang, Khanikaev, Hazefi, Rechtsman, Chen, Joannopoulos, Hassani1}. The ideas underlying unidirectional photonic transport were initially developed for electrons in crystalline materials (electronic topological insulators) \cite{Kane, Hasan}. In particular, the most common approach \cite{Haldane,Haldane2} consists in utilizing 2D photonic crystals with degenerate Dirac cones in their photonic band structure. The degeneracy can be lifted by breaking time-reversal symmetry, which opens a band gap and leads to topologically non-trivial photonic bands with non-zero Chern numbers. The unidirectional mode can then propagate at the interface between a PTI and an ordinary photonic insulator or simple opaque material. Time-reversal symmetry can be lifted by applying a static magnetic field to a gyromagnetic material \cite{Wang,Poo}, time-harmonic modulation of coupled dielectric resonators \cite{Fang,Hazefi}, or by translating the modulation from time-domain to spatial domain \cite{Rechtsman}. Alternatively, photonic analogues of time-reversal-invariant topologically protected $Z_2$ states have been proposed \cite{Khanikaev,Chen,Wu}. In this case, photons are separated into different 'spin' sub-spaces based on their polarization and ‘spin-orbit’ coupling is introduced through exploiting such non-trivial characteristics of metamaterials as chirality, bi-anisotropy and hyperbolicity. 

In the above cases one deals with periodic media, and the underlying wave vector space is isomorphic to a torus \cite{Hughes}. However, the possibility of continuous material to support topologically protected unidirectional surface plasmon polaritons has been also recently been shown \cite{Arthur, Gao, Yang, Mario2}. In particular, immune to backscattering edge/surface modes were first predicted at the interface between biased magneto-plasma material and metal \cite{Arthur}. In \cite{Gao, Yang} the authors consider Chern number calculations, but based on equi-frequency surfaces (EFS). The formalism that allows for the formal extension of the concept of PTI to the case of continuous media based on global features of the eigenmodes has been developed in Ref. \cite{Mario2}. In this case, the Chern number is calculated for a set of eigenmodes of the continuous material (rather than for the photonic bands), with the wavenumber taking values in the range from $-\infty$ to $\infty$. The Chern number computed in such a way has analogous meaning to that for electronic or photonic bands in a periodic structure. In particular, at the interface between a PTI and an ordinary insulator, the number of one-way states is equal to the difference of gap Chern numbers between the PTI and the ordinary insulator, where the gap Chern number $ (C_{gap} = \sum_i C_i ) $ is the sum of the Chern numbers of all modes below the band gap \cite{Hatsugai,scott,scott2,Gao}. If one of the materials is topologically trivial ($  C_{gap} = 0 $) like metal or air, the other gap Chern number determines the number of topologically protected one-way interface states.

The previous works on PTIs were focused primarily on 2D structures and demonstrated immunity of topologically protected one-way interface modes with respect to backscattering by 2D defects. However, any realistic waveguides and defects are necessarily 3D. This fact may cause several problems with the practical implementation of PTIs. First of all, scattering of the interface mode at a 3D defect is not limited to backscattering. In fact, side scattering may be significant and PTIs seem not to offer any protection against this. Moreover, modes of the finite waveguides are cavity modes rather than modes of the infinite medium. Thus, the finite waveguide might have modes with frequencies in the bandgap region of bulk PTIs, which may lead to coupling of interface modes to the bulk cavity modes, and thus to significant decrease of the one-way mode propagation distance. In this work we consider these issues, taking the interface between a biased magneto-plasma and metal as our model system. The magneto-plasma was considered before in various contexts, such as sub-diffractional imaging \cite{Zhang} and magnetic field induced transparency \cite{Gad}. The topological protection of the interface mode in magneto-plasma was demonstrated \cite{Arthur}, although neither formal assessment of the topological bandstructure was made nor were the Chern numbers calculated (the work in \cite{Gao, Yang} is highly related, but is based on EFS-related calculations).

In this paper, we analytically calculate Chern numbers for the eigenmodes supported by a biased magneto-plasma based on global dispersion behavior of the eigenmodes of the bulk structure, using the continuum material theory developed in \cite{Mario2}. (see Supplemental Information for details). We demonstrate that magneto-plasma material possesses topologically non-trivial bulk electromagnetic modes with non-zero Chern number. This leads to propagation of the topologically non-trivial surface modes at the interface between the magneto-plasma and metal that are immune to backscattering. We consider interaction of these modes with 3D effects and demonstrate that side-scattering on these defects significantly decreases surface propagation lengths and thus has to be dealt with in any realistic waveguide. We also predict that radiative losses due to the finite sizes of the waveguide are important. We offer two different waveguide geometries that allows to counter these effects and achieve lossless propagation of surface modes immune to back-scattering on defects and waveguide inhomogeneities.

\section*{Results}

\subsection*{Chern number and bulk-edge correspondence}

The ideas underlying electronic topological insulators can be extended to the photonic case by utilizing the correspondence between Maxwell's equations and Schr\"{o}dinger's equation \cite{Haldane, Haldane2, Mario2}. We assume lossless dispersive materials \cite{Haldane, Haldane2}, such that the continuum material can be characterized by dimensionless real-valued parameters $\overline{\epsilon },~\overline{\mu },~\overline{\xi },~\overline{\varsigma }$, representing permittivity, permeability and magneto-electric coupling tensors. By defining matrices 
\begin{align}
	&\boldsymbol{M} =\left( 
	\begin{array}{cc}
		\epsilon _{0}\overline{\epsilon } & \frac{1}{c}\overline{\xi } \\ 
		\frac{1}{c}\overline{\varsigma } & \mu _{0}\overline{\mu }%
	\end{array}%
	\right) ,~~~\hat{N}=\left( 
	\begin{array}{cc}
		0 & i\nabla \times \mathbf{I}_{3\times 3} \\ 
		-i\nabla \times \mathbf{I}_{3\times 3} & 0%
	\end{array}%
	\right) ~  \label{mat}
\end{align}
where $ \boldsymbol{M} $ is Hermitian and real-valued, we can write Maxwell's equations in compact form \cite{Mario1},
\begin{equation}
	\left( 
	\begin{array}{cc}
		0 & -\mathbf{k}\times \mathbf{I}_{3\times 3} \\ 
		\mathbf{k}\times \mathbf{I}_{3\times 3} & 0%
	\end{array}%
	\right) \cdot \left( 
	\begin{array}{c}
		\mathbf{E} \\ 
		\mathbf{H}%
	\end{array}%
	\right) =\left( 
	\begin{array}{cc}
		\omega \epsilon _{0}\overline{\epsilon } & 0 \\ 
		0 & \omega \mu _{0}\mathbf{I}_{3\times 3}%
	\end{array}%
	\right) \cdot \left( 
	\begin{array}{c}
		\mathbf{E} \\ 
		\mathbf{H}%
	\end{array}%
	\right) .  \label{EE4}
\end{equation}
In the above, $ \omega$ is the radian frequency and $ \boldsymbol{\mathrm{k}} = -i \nabla $ is the wave vector, and in the following we will set the magneto-electric coupling tensors to zero. If we define $ \boldsymbol{f}=\left[ \boldsymbol{\mathrm{E}}, \boldsymbol{\mathrm{H}}  \right]^{\mathrm{T}} $ then Maxwell's equations become 
\begin{equation}
	\hat{H}_{cl}\cdot \boldsymbol{f}_{n}=E_{n} \boldsymbol{f}_{n}  \label{PEE1}
\end{equation}%
where $\hat{H}_{cl} = \boldsymbol{M}^{-1} \cdot \hat{N}$ plays the role of a classical Hamiltonian. 

The Berry potential $ \mathbf{A}_{n}=i\left\langle \boldsymbol{f}_{n}|\nabla _{\mathbf{k}}\boldsymbol{f}_{n}\right\rangle $ for the case of dispersive materials, defined by $ M{(\omega)} $, is \cite{Haldane} 
\begin{equation}
\mathbf{A}_{nk}=\text{Re}\{i\boldsymbol{f}_{nk}^{\ast }\cdot \frac{\partial }{%
		\partial \omega }(\omega \boldsymbol{M}{(\omega )}) \cdot \partial _{k}\boldsymbol{f}_{n,k}\},
\label{Eq:60}
\end{equation}
and the Chern number can then be calculated as \cite{Mario2} (see also SI for details)
\begin{equation}
C_{n}=\lim\limits_{k\rightarrow \infty }(A_{n,\phi =0}k)-\lim\limits_{k\rightarrow 0^{+}}(A_{n,\phi =0}k) \label{Eq:Chern},
\end{equation}
where $ A_{nk}=\mathbf{A}_{nk}\cdot \hat{\mathbf{\phi}}$. 

\begin{figure}[ht]
	\begin{center}
		\noindent
		\includegraphics[width=6in]{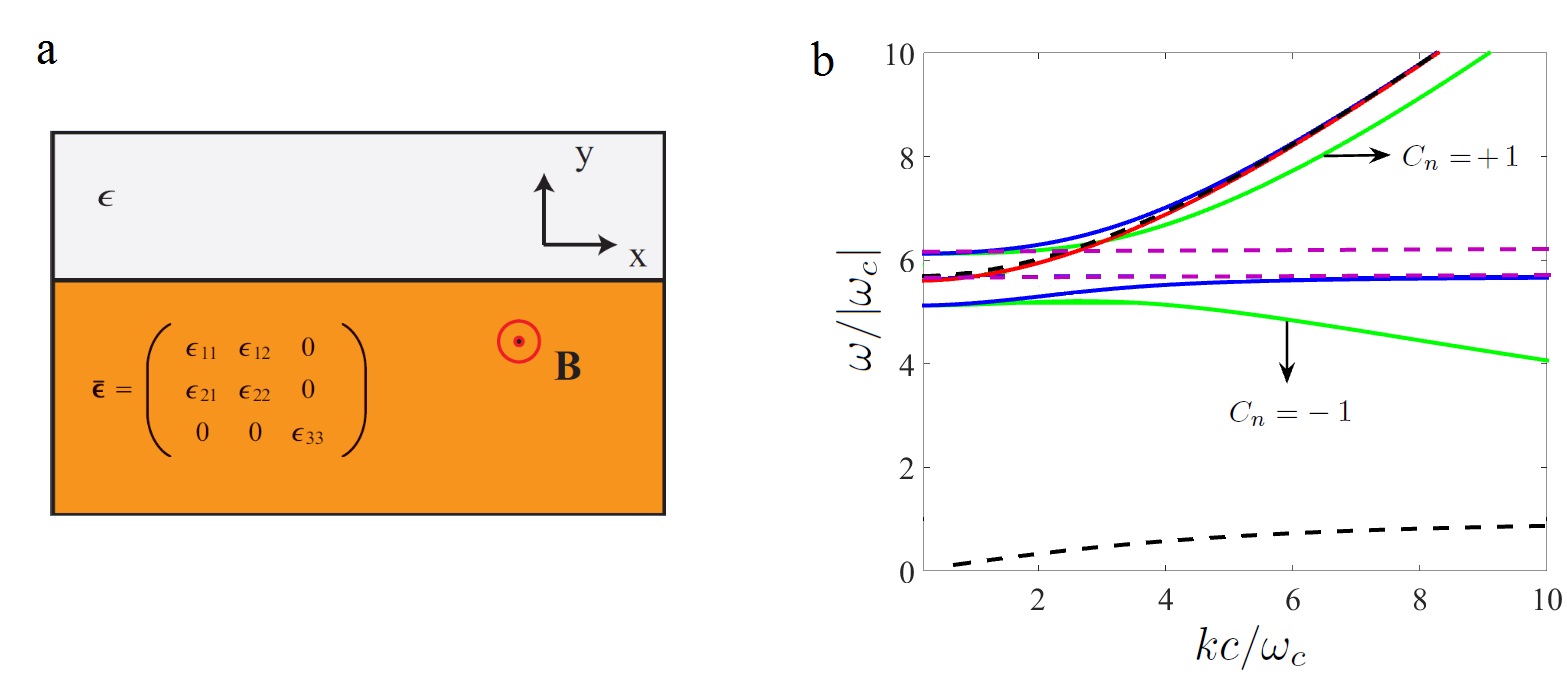}
		\caption{a. Interface between a magnetic-field biased plasma (bottom) and a simple scalar material (top). b. TM and TE band diagram for $ \omega_p / 2\pi=9.7 $ THz and $ \omega_c/ 2 \pi= 1.73 $ THz; $ \omega_p = 5.6 \omega_c $. Solid blue: bulk TM mode for local model, solid red: bulk TE mode for local model, solid green: bulk TM mode for nonlocal model with spacial cutoff $ k_{max} = 10|\omega_c|/c $, dashed black: SPP mode dispersion at the interface of magneto-plasma and metal (local model). The dashed purple lines show the bulk TM mode gap (local modal).}\label{Fig2}
	\end{center}
\end{figure}

Let us consider a magnetized plasma in the Voigt configuration and assume interface mode propagation in the direction perpendicular to the bias magnetic field $\mathbf{B}$), as depicted in Fig. \ref{Fig2}a.  For a single-component plasma biased with a static magnetic field $\mathbf{B} =\mathbf{z}B_{z}$, the permeability is $\mu=\mu_0$ and the relative permittivity has the form of a Hermitian antisymmetric tensor,
\begin{equation}\label{moe}
	\overline{\epsilon }=\left( 
	\begin{array}{ccc}
		\epsilon _{11} & \epsilon _{12} & 0 \\ 
		\epsilon _{21} & \epsilon _{22} & 0 \\ 
		0 & 0 & \epsilon _{33}%
	\end{array}%
	\right) 
\end{equation}%
where
\begin{align}
	\varepsilon _{11}& =\varepsilon _{22}=1-\frac{\omega _{p}^{2}}{\omega
		^{2}-\omega _{c}^{2}}\text{,  \: }\varepsilon _{33}=1-\frac{\omega _{p}^{2}}{%
		\omega ^{2}}\text{,  \: }\varepsilon _{12} =-\varepsilon _{21}=i\frac{-\omega _{c}\omega _{p}^{2}}{%
		\omega \left( \omega ^{2}-\omega _{c}^{2}\right) }, \label{BMPM}
\end{align}%
and where the cyclotron frequency is $\omega _{c}=\left( q_{e}/m_{e}\right)
B_{z}\ $and the plasma frequency is $\omega
_{p}^{2}=N_{e}q_{e}^{2}/\varepsilon _{0}m_{e}$. In the above, $N_{e}$ is the free electron density, and $q_{e}$ and $m_{e}$ are the electron charge and mass, respectively. 

Let us first consider photonic modes of the infinite magnetized plasma. We assume electromagnetic wave propagation in the $xoy$ plane, $ \mathbf{k} = (k_x,k_y,0) $. In this case the transverse electric (TE) ($ E_z \neq 0,~ H_z=0 $) and transverse magnetic (TM) ($ E_z = 0,~ H_z \neq 0 $) modes are decoupled and their dispersion is given by

\begin{align}
	&k^2 = \frac{\epsilon_{11}^2 + \epsilon_{12}^2 }{\epsilon_{11}} \left(\frac{\omega}{c}\right)^2 ~~ (\mathrm{TM~modes}) \text{,  \: } k^2 = \epsilon_{33} \left(\frac{\omega}{c}\right)^2 ~~ (\mathrm{TE~modes}).
\end{align}
Figure \ref{Fig2} depicts the band diagram of the bulk TM (solid blue) and bulk TE (solid red) polarized modes for the case $ \omega_p = 5.6 \omega_c $. The dispersion of the TE mode shows a band gap for $ \omega < \omega_p $ and the TM modes are organized into two branches separated by a band gap. 

Using Eq. \eqref{Eq:Chern} (see SI for details) we calculate the Chern numbers for all bands. The Chern number for the TE band is trivial, $ C_n=0 $, and  the Chern number of the high-frequency TM band is 1. However, the Chern number of the low frequency TM band is not an integer, $C = -sgn(\omega_c) / \sqrt{1+ (\omega_p / \omega_c)^2} -1$. This is associated with the Hamiltonian of the medium not being sufficiently well behaved as $ k \rightarrow \infty $ for the local model \cite{Mario2}. 

An integer Chern number for both branches can be obtained using a non-local material response with a high-wavenumber spatial cutoff \cite{Mario2} $ \boldsymbol{M}_{reg}(\omega, \boldsymbol{k} ) = \boldsymbol{M}_{\infty} + \frac{1}{1+k^2/k_\mathrm{max}^2} \left \{ \boldsymbol{M}({\omega}) - \boldsymbol{M}_{\infty} \right\} $ where $ \boldsymbol{M}_{\infty} = \lim\limits_{\omega \rightarrow \infty} \boldsymbol{M}({\omega}) $. For this nonlocal material response, the band diagram and Chern numbers have been obtained using $ k_\mathrm{max}= 10 |\omega_c|/c $ (for TM modes these are shown as solid green lines, and the Chern number calculation is detailed in the Supplemental Information). As can be seen in Fig. \ref{Fig2}, for the lower TM mode the lower frequency limit of the band gap in the presence of spacial cutoff changes somewhat from the local case (blue lines), and we have integer Chern numbers for all modes, such that $ C_{gap} = -1 $, which indicates the presence of one back-scattering-protected SPP.  From (27) and (28) in the supplemental information (SI), the lower and upper frequency of the band gap (of the TM modes for the infinitely-wide case) is $ \omega_L/\omega_c=5.66 $ and $ \omega_H/\omega_c=6.07 $. 

Let us now consider surface plasmon polaritons (SPP) propagating at the interface between magnetized plasma and isotropic non-magnetic material. The dispersion equation of the TM-polarized SPP mode (assuming local model for dielectric response of magnetized plasma) is \cite{Arthur}
\begin{equation}
\frac{\gamma_m}{\epsilon_m} + \frac{\gamma_\nu}{\epsilon_{eff}} = i \frac{\epsilon_{12} k_x}{\epsilon_{11} \epsilon_{eff} }
\end{equation}
where $ k_x $ is the propagation constant of the interface state, $ \epsilon_m $ is the permittivity of the metal, $ \epsilon_{11} $ and  $ \epsilon_{12} $ describe the permittivity of the magneto-plasma material (\ref{BMPM}), $ \epsilon_{eff} = \frac{\epsilon_{11}^2 + \epsilon_{12}^2 }{\epsilon_{11}}  $, $ \gamma_m  = \sqrt{k_x^2 - (\omega / c)^2 \epsilon_m}$ and $ \gamma_{\nu}  = \sqrt{k_x^2 - (\omega / c)^2 \epsilon_{eff}}$. In the following results we consider silver as the metal, which, at $\omega/2\pi=10 $ THz $(\omega / \omega_c=5.78)$, has permittivity $ \epsilon = (-4.00-i1.69)\times10^4 $. In order to focus on the important physics, we first consider lossless silver ($ \epsilon = -4.00\times10^4 $; the results using this value are not too different than for a perfect-electric conductor). The dashed black lines in Fig. \ref{Fig2} show dispersion of the SPP (edge mode). The bandgap is indicated by the dashed purple lines (for the local model). As can be seen, the upper SPP dispersion line crosses the bandgap of the bulk modes, which is the frequency range of interest in the following. 

\subsection*{Full-wave simulation of one-way interface mode propagation}

Certainly, the most celebrated property of topological materials is their potential to support propagation of backscattering-protected modes at the interface between two topologically distinct materials. Furthermore, one wants to operate in the band gap of both materials (or in the bandgap of the PTI and in an opaque regime for the trivial material), to avoid losses due to radiation and diffraction at surface discontinuities. In this section we consider one way propagation of the surface mode at the interface between the biased magnetoplasma and metal. All numerical simulations are performed using CST Microwave Studio.

\begin{figure}[h!]
	\begin{center}
		\noindent
		\includegraphics[width=5in]{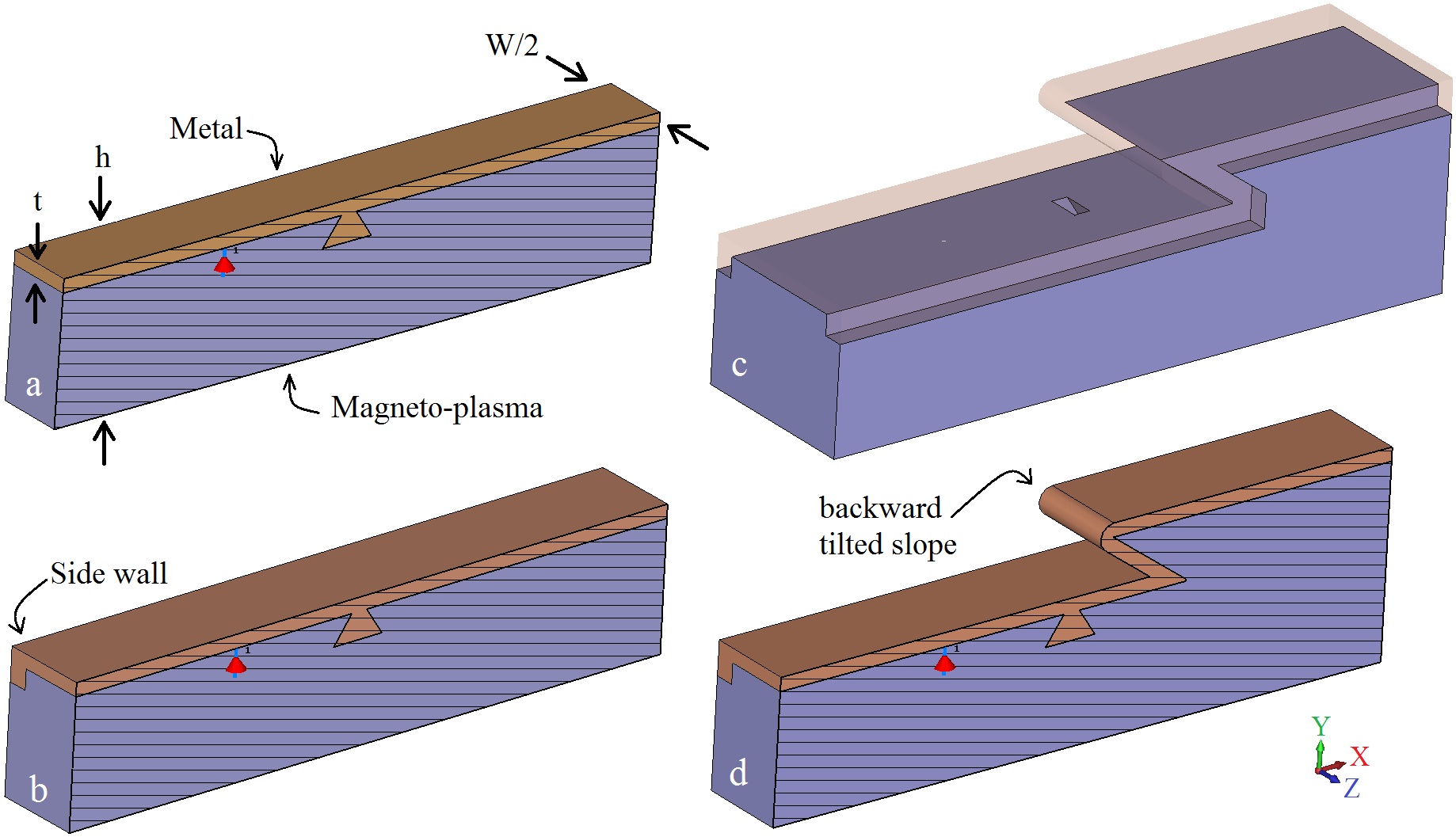}
		\caption{a- Perspective cross section view of the interface between a metal coating (top brown material) and a magneto-plasma material (lower purple material), including a metallic pyramidal barrier near a dipole source (red arrow, vertically-polarized). The width of the interface (along $ z $) is $ W $ and the coating thickness is $t$ . b- Perspective cross section view of the same structure with partial side walls. c- Perspective view of a plasmonic waveguide incorporating an inclined step discontinuity. d- Perspective cross section view of the waveguide with inclined step discontinuity. }\label{st1}
	\end{center}
\end{figure}

Previous works on the subject of PTI focused on 2D waveguiding structures and studied propagation of the topologically protected modes at the interface between two semi-infinite half-spaces filled with topologically distinct materials. The 2D geometry however, can not account for the presence of 3D defects, which is a more realistic scenario. In the case of a topologically protected mode interacting with such a defect, the energy can be scattered sideways (upwards/downwards scattering/radiation is assumed suppressed due to being in the bandgap of the bulk materials). It is obvious that topological protection prohibits backward scattering of the mode. However, sideways scattering can still be an issue as it is not subjected to this protection. Thus, it is important to estimate how significant this decay channel is. To address this issue, we study excitation of the TM mode at the interface between magnetized plasma and metal for several waveguiding geometries (see Fig. \ref{st1}). We choose an electrical dipole oriented normal to the interface as the excitation source (red arrow in Fig. \ref{st1}), as it couples predominantly to the TM modes. In what follows we assume that the dipole frequency is in the band gap of the magneto-plasma material and is equal to $ \omega / 2 \pi=10 $ THz ($ \omega / \omega_c=5.78 $ THz).

The first waveguiding geometry (cross-section is shown in Fig. \ref{st1}a ) consists of a thin metal layer (thickness $t$, width $W$) placed on top of a magneto-plasma material (thickness $h-t$, width $W$). We added a metallic pyramid at the interface to serve as a 3D defect. The size of this defect is comparable to the SPP wavelength, such that the path along the contour of the defect is approximately a half wavelength of the SPP ($\lambda_{SPP} \approx 162~ \mu $m). The waveguide is $480~ \mu$m long (x-direction), the right end is at $x=350~ \mu$m, and the left end is at $x=-130~ \mu$m); the source is at $x=0$ centered across the width of the structure ($z=0$). The ends are terminated by an open circuit (vacuum). In the following, the width of the interface (along $ z $) is $ W=100~\mu $m, the coating thickness is $t=10 ~ \mu$m, and we assume lossless or lossy silver as the metal. For the structure with partial sidewalls, the depth of the side wall is $ 15~ \mu $m.

\begin{figure}[h!]
	\begin{center}
		\noindent
		\includegraphics[width=6in]{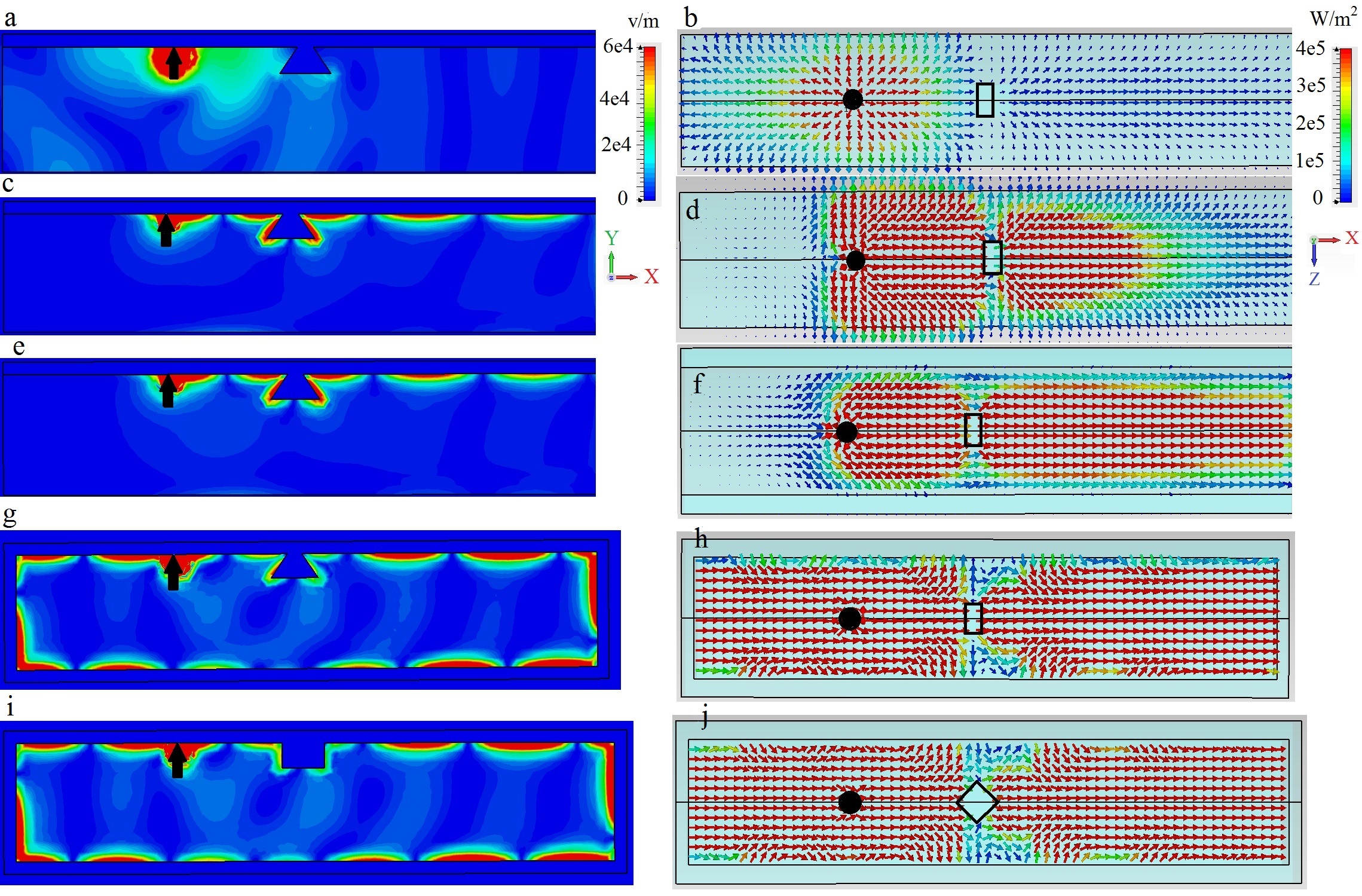}
		\caption{Electric field distribution for the surface mode excited by electric dipole (black arrow and circle) at the interface between metal and magneto-plasma material ($\omega_p / 2 \pi=9.7 $ THz). Left panels: distribution of the electric field intensity in the $x-y$ plane at the cross-section $z = 0$ (see Fig. \ref{st1} for details). Right panels: Top view of power distribution at the interface. (a,b) Non-biased ($ \omega_c=0 $) magneto-plasma material (see Fig. \ref{st1}-a for geometry). (c,d). Biased ($ \omega_c / 2 \pi=1.73 $ THz) magneto-plasma material (see Fig. \ref{st1}-a for geometry). (e,f). Biased magneto-plasma material with partial side wall (see Fig. \ref{st1}-b for geometry). (g,h). Biased magneto-plasma material with metal layer surrounding all faces. (i,j). Same as in panels (g,h) but for a different 3D defect. }\label{elec1}
	\end{center}
\end{figure}

First, we consider excitation of the surface wave at the interface between lossless silver and non-biased magneto-plasma ($ \omega_c=0 $). Distributions of the electric field and power carried by the wave are presented in Figs. \ref{elec1}a,b, respectively. This is the case when time reversal symmetry is preserved and thus, as one can see from Figs. \ref{elec1}a,b, the excitation propagates in all directions. Moreover, backscattering is significant and thus the surface wave is mostly reflected by the defect. In order to address this issue we break non-reciprocity by applying magnetic bias, $ \omega_c/2 \pi=1.73 $ THz. Lifting the time-reversal symmetry leads to appearance of the topologically non-trivial photonic modes in the magneto-plasma (see Fig. \ref{Fig2}) and thus to the unidirectional surface modes at the interface (Figs. \ref{elec1}c,d). One can clearly see unidirectional propagation of the mode from the power distribution presented in Fig. \ref{elec1}d. As backscattering is suppressed the excitation can efficiently propagate around the defect.

Somewhat puzzling is that the propagation distance of the surface wave is rather low despite the fact that both magneto-plasma and silver are assumed to be lossless and backscattering is suppressed. This is due to the radiative damping of the surface polariton. As one can see from the power distribution in Fig. \ref{elec1}d, the surface wave carries power not only in the waveguiding direction ($x$ in this case), but also in the orthogonal $z$ direction. The reason is that both the dipole source and defect can excite waves propagating in all directions, while topological protection prevents propagation in the negative $x$ direction only, and not in the $z$ direction. As the waveguide is finite in the $z$ direction, the power carried in that direction is radiated into free-space at the waveguide edges, leading to radiative damping of the surface wave. The radiative losses due to the excitation by the dipole can however be eliminated by choosing a source that excites only modes propagating along the $x$ direction (a planar Yagi-Uda or waveguide source, for example). However, the scattering by the defect is independent of the choice of the source and is inherently linked to possibe 3D waveguide imperfections, and thus has to be addressed in a different way. 

In order to solve the problem of multi-directional wave scattering by the defect and provide efficient channeling of the power along a given direction, we choose to protect the wave propagation by extending the metal cover layer down the sides of the waveguide  (see Fig. \ref{st1}b). As can be seen from Figs. \ref{elec1}e,f this leads to strong confinement of the surface wave in the $z$-direction, eliminating radiative losses due to waves propagating in $z$-direction, and thus to the increase of the surface polariton propagation distance. Nevertheless, we can still observe that the surface plasmon experiences damping, which should be prevented by the fact that the plasmon frequency is in the bandgap of the materials comprising the waveguide (Fig. \ref{elec1}-b, c) and both materials are lossless. However, it should be pointed out that the indicated bandgap is for infinite materials where we take into consideration bulk modes only. Our waveguide is finite both in $y$ and $z$ directions (see Fig. \ref{st1}) and thus the waveguide eigenmodes are cavity modes rather than bulk modes of infinite the materials. Thus, a few of the cavity modes can occur in the bandgap region and the surface plasmon can couple to these modes, leaking energy through the bottom of the waveguide and the portion of the waveguide sides not covered by metal. This is supported by careful examination of Figs. \ref{elec1}c,e,g,i, where electric field penetration inside the magneto-plasma material can be clearly seen. 

Power loss into bulk modes can be overcome by enclosing all of the faces of the waveguide by a metal layer (here, chosen to be 10 $\mu$m thick). We can see from Fig. \ref{elec1}g,h that the plasmon circulates the entire structure in a clockwise manner, without any losses. There are, in fact, coupling of energy to the bulk modes, but this is eventually returned back to the surface mode (some energy also circulates in the $y-z$ plane, around the circumferential of the structure, due to side scattering). By opening the end-faces of the structure (planes perpendicular to $x$ direction) we can let energy out, thus implementing a unidirectional waveguide without losses. Reversing the direction of the magnetic field results in counter-clockwise rotation of the plasmon. Fig. \ref{elec1}-i and j show similar results to panels g and h, except for a different shape 3D defect. For \ref{elec1}g, videos showing clockwise and counterclockwise rotation of the SPP for positive and negative bias, respectively, are available in the Supplemental Material

\begin{figure}[h!]
	\begin{center}
		\noindent
		\includegraphics[width=6in]{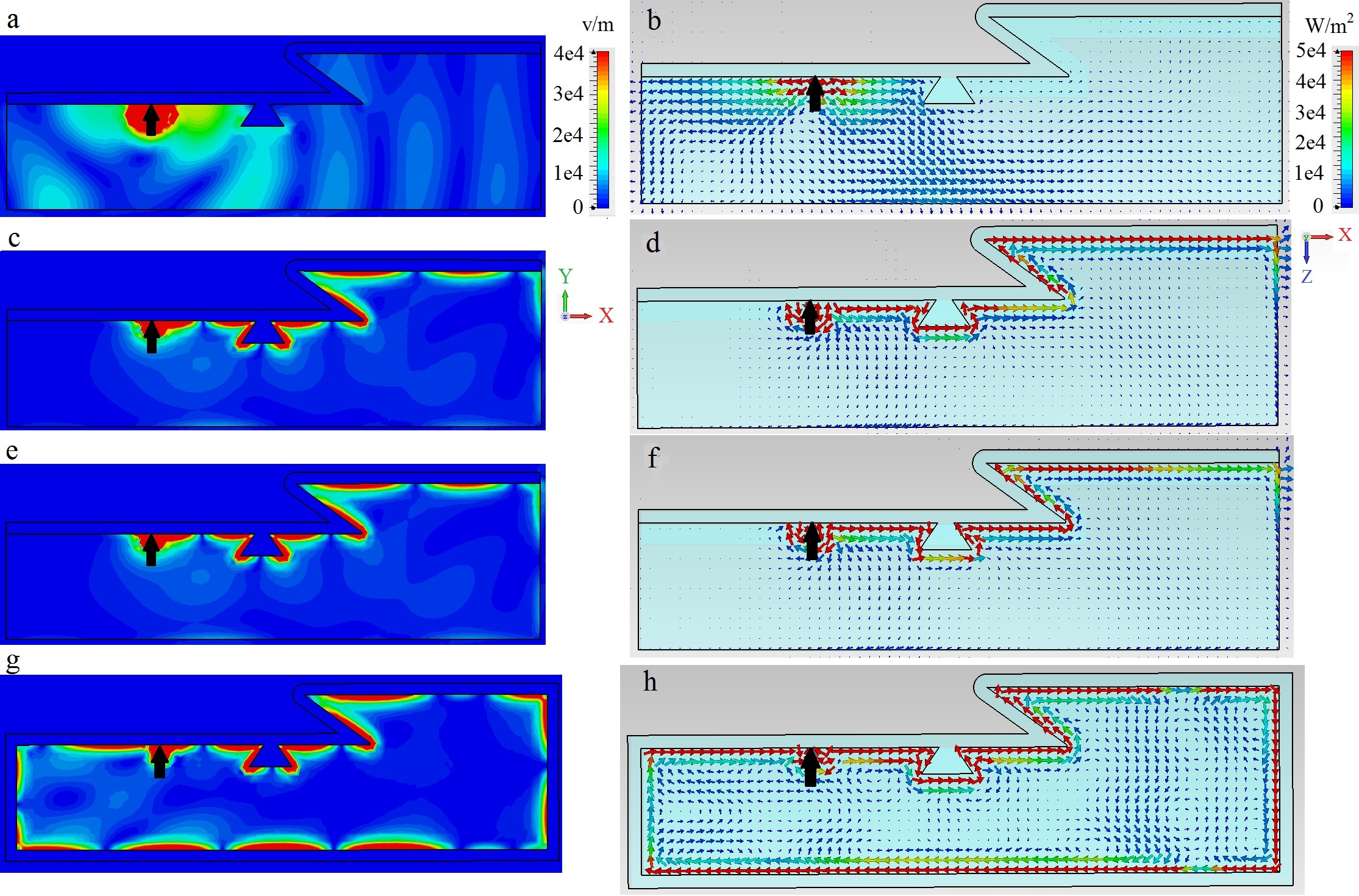}
		\caption{Influence of pronounced waveguide discontinuity on the propagation of the surface wave excited by an electric dipole (black arrow) at the interface between metal and magneto-plasma material ($\omega_p / 2 \pi=9.7 $ THz). Unless specified otherwise, the waveguide sides are partly covered by metal (see Fig. \ref{st1}c,d for details). Left panels: distribution of the electric field intensity in the $x-y$ plane at the cross-section $z = 0$ (see Fig. \ref{st1}c,d for details). Right panels: Corresponding power distribution. (a,b) Non-biased ($ \omega_c=0 $) magneto-plasma material and lossless silver. (c,d). Biased ($ \omega_c / 2 \pi=1.73 $ THz) magneto-plasma material and lossless silver. (e,f). Biased magneto-plasma material and lossy silver. (g,h). Biased magneto-plasma material with layer of lossless silver surrounding all faces.  }\label{elec2}
	\end{center}
\end{figure}

As further demonstration of the uni-directional nature of this waveguiding structure, we consider an waveguide with both a defect and an inclined step discontinuity, as depicted in Fig. \ref{st1}c,d. Partial side walls are again used to prevent field leakage into space. The length of the tilted slope is comparable to the SPP wavelength.  Figure \ref{elec2} shows a side view of the electric field and power distribution for non-biased (a,b) and biased (c,d) magneto-plasma with lossless silver, for biased magnetoplasma with lossy silver (e,f), and for biased magneto-plasma with a lossless silver layer surrounding all faces (g,h). For the case of non-biased magneto-plasma (a,b), the SPP at the interface propagates in both directions, and can not penetrate through the waveguide discontinuity. As in the case presented in Fig. \ref{elec1}, magnetic biasing (\ref{elec2}c,d) leads to the unidirectional surface wave propagation at the interface, which is insensitive to the presence of the waveguide discontinuity. However, radiative damping due to the coupling between surface and bulk modes is still substantial. As a next step, we study the effect of the losses in silver on surface wave propagation. However, in the low THz frequency range losses are relatively small and silver behaves similar to PEC. Thus, losses do not affect too significantly the surface wave propagation, as can be seen from Fig. \ref{elec2}-e,f. Finally, in order to address the issue of radiative losses due to the surface-bulk coupling, we enclosed the magneto-plasma by a 10 $\mu$m thick lossless metal layer (\ref{elec2}-g,h). As with Fig. \ref{elec1}-g,i, we observe clockwise circulation of the surface wave around the structure, without any indication of damping.

\begin{figure}[h!]
	\begin{center}
		\noindent
		\includegraphics[width=5.5in]{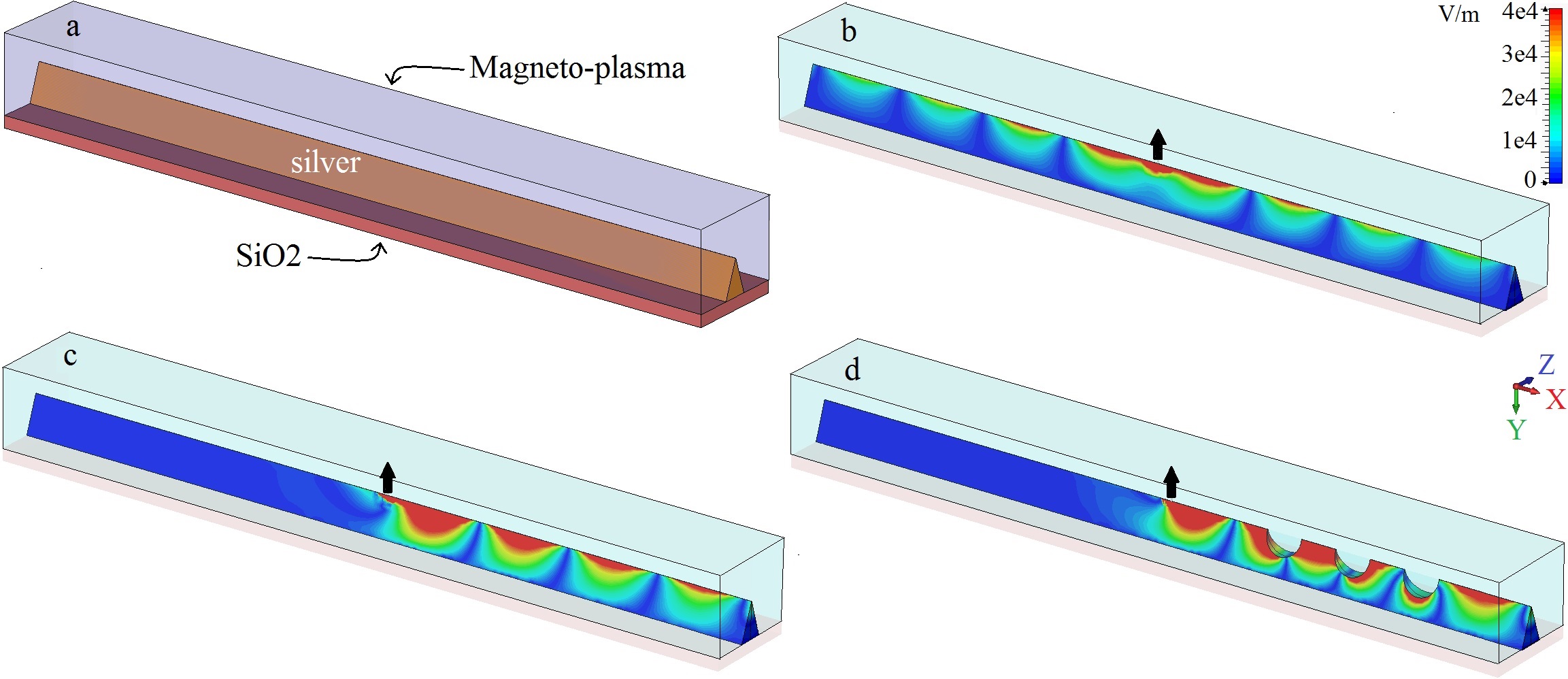}
		\caption{a- Perspective view of the silver ridge waveguide with height $ 30~\mu $m and opening angle $ 20 $ degree mounted on $ \mathrm{SiO_2} $ substrate and covered by magneto-plasma material. b- Electric field distribution on the ridge for non-biased ($ \omega_c=0 $) case and c- for biased ($ \omega_c / 2 \pi=1.73 $ THz) case. d- shows the biased case with defects on the ridge. The dipole source is indicated by a black arrow, lossy silver is considered, and in all cases the excitation frequency $\omega/2\pi = 10$ THz. $ \omega_p / 2 \pi=9.7 $ THz.}\label{ridge}
	\end{center}
\end{figure}

As a possible alternative waveguide that eliminates side-scattering, we consider a ridge waveguide, depicted in Fig. \ref{ridge}-a. There is an intense effort to channelize SPP propagation in photonic integrated circuits \cite{Raether,William,Atwater,Jianwei,Hassani2}, and ridge waveguides are one of the promising structures to fulfill this demand \cite{Jianwei}. In this part we demonstrate a one-way plasmonic ridge waveguide as shown in Fig. \ref{ridge}-a. The ridge is made of lossy silver with height $ 30~\mu $m and opening angle $ 20 $ degree placed on the top of a substrate made of $ \mathrm{SiO_2} $ and covered by magneto-plasma material. The magneto-plasma material is biased along the $ z $ direction. A y-polarized dipole source is placed at the top of the ridge and radiates at $ 10 $ THz. Fig. \ref{ridge}-b shows the absolute value of the electric field on the ridge covered by non-biased $ (\omega_c / 2 \pi = 0) $ magneto-plasma. As can be seen, we have both right and left SPP propagation. But, by turning on the bias $ (\omega_c / 2 \pi = 1.73 ~ \mathrm{THz}) $, as shown in Fig. \ref{ridge}-c, we have SPP propagation only toward right without any backward propagation. Although it is not shown here, in all considered waveguiding structures, by changing the direction of bias toward $-z$, we obtain only leftward SPP propagation. Fig. \ref{ridge}-d shows the one-way propagation of the SPP in the presence of defects in the ridge waveguide. As expected, the defect does not scatter energy, and the SPP propagates around the defect.

All the analysis assumed the Voight configuration, where propagation is perpendicular to the biasing field. Existence of such a mode is valid for the case of in-plane bends. If we have a bend in the plane of the biasing field, then the propagation in the bend region is not normal to the biasing field (unless the bias direction changes with the bend). For propagation perpendicular to the bias, modes are decoupled TE and TM, whereas for propagation parallel to the bias, modes are right- and left-handed circularly polarized, which will lead to back scattering in the bend region. 

Concluding, in this paper we studied the effects of 3D defects and the finite size of realistic waveguides on propagation of topologically protected unidirectional surface modes. We calculated analytically the Chern number of the bulk electromagnetic modes in a continuum biased magneto-plasma, and demonstrated the topologically non-trivial character of these modes, which leads to appearance of topologically protected unidirectional surface waves at the interface between the magneto-plasma and plasmonic metal. While it is well known that suppressed backscattering allows these modes to carry energy around 2D defects in the plane of interface without any losses, we demonstrated that 3D defects can have a substantial detrimental effect on unidirectional surface mode propagation length, due to the fact that 3D defects can scatter energy sideways. While topological protection prevents surface mode energy from flowing backwards, no protection is provided for energy flowing sideways. The energy leaking into free space from the sides of waveguides can lead to significant radiation losses. Another mechanism of losses is coupling of surface modes to the cavity bulk modes of a realistic finite-size waveguide. Even though bulk modes in magneto-plasma have a bandgap at the frequency of waveguide operation, cavity modes can exist in the band gap. This is an inevitable consequence of the finiteness of any realistic waveguide structure. We suggested two different ways to solve the problems of side scattering and cavity modes. One consists of enclosing the magneto-plasma material with metal, while another relies on using a ridge waveguide.  


\section*{Additional Information}

The authors declare no competing financial interests.

\section*{Author contributions statement}

All authors conceived of the work and reviewed the manuscript. AH performed the simulations and wrote the initial manuscript draft.

\end{document}


\title{Supplemental material for \\ \textbf{The effects of three-dimensional defects on one-way surface plasmon propagation for photonic topological insulators comprised of continuous media}}
\author{S. Ali Hassani Gangaraj, Andrei Nemilentsau, George W. Hanson  \\ \\ Department of Electrical Engineering, University of Wisconsin-Milwaukee \\ Milwaukee Wisconsin 53211, USA \\ \\ Email: Hassani4@uwm.edu }

\maketitle

\section{Maxwell's equation-Schr\"{o}dinger equation duality}

We first assume lossless and dispersionless materials characterized by dimensionless real-valued parameter $\overline{\epsilon },~\overline{\mu },~\overline{\xi },~\overline{\varsigma }$, representing permittivity, permeability and magneto-electric
coupling tensors. In such a medium the Maxwell's equations (considering time convention $ e^{-i\omega t} $) are

\begin{align}
& \nabla \times \mathbf{E}=-\mu _{0}\overline{\mu }\cdot \frac{\partial 
	\mathbf{H}}{\partial t}-\frac{\overline{\varsigma }}{c}\cdot \frac{\partial 
	\mathbf{E}}{\partial t}-\mathbf{J}_{m}  \notag  \label{Eq:17} \\
& \nabla \times \mathbf{H}=\epsilon _{0}\overline{\epsilon }\cdot \frac{%
	\partial \mathbf{E}}{\partial t}+\frac{\overline{\xi }}{c}\cdot \frac{%
	\partial \mathbf{H}}{\partial t}+\mathbf{J}_{e}.
\end{align}%
By defining the matrices%
\begin{align}
\boldsymbol{M}& =\left( 
\begin{array}{cc}
\epsilon _{0}\overline{\epsilon } & \frac{1}{c}\overline{\xi } \\ 
\frac{1}{c}\overline{\varsigma } & \mu _{0}\overline{\mu }%
\end{array}%
\right) ,~\hat{N}=\left( 
\begin{array}{cc}
0 & i\nabla \times \mathbf{I}_{3\times 3} \\ 
-i\nabla \times \mathbf{I}_{3\times 3} & 0%
\end{array}%
\right) ,~  \label{mat} \\[5pt]
\boldsymbol{f}& =\left( 
\begin{array}{c}
\mathbf{E} \\ 
\mathbf{H}%
\end{array}%
\right) ,~\boldsymbol{g}=\left( 
\begin{array}{c}
\mathbf{D} \\ 
\mathbf{B}%
\end{array}%
\right) =\boldsymbol{M} \cdot \boldsymbol{f},~\boldsymbol{J}=\left( 
\begin{array}{c}
\mathbf{J}_{e} \\ 
\mathbf{J}_{m} \notag
\end{array}%
\right) \ 
\end{align}%
where $ \boldsymbol{M} $ is Hermitian and real-valued, we can write Maxwell's equations in a
compact form \cite{Mario1},%
\begin{equation}
\hat{N}\cdot \boldsymbol{f}=i\left[ \frac{\partial \boldsymbol{g}}{\partial t}+\boldsymbol{J}\right] =i\left[ \boldsymbol{M}\frac{%
	\partial \boldsymbol{f}}{\partial t}+\frac{\partial \boldsymbol{M}}{\partial t}\boldsymbol{f}+\boldsymbol{J}\right] . \label{ME2}
\end{equation}%
Note that the units of the sub-blocks of $ \boldsymbol{M} $ differ (as do the dimensions of the 6-vectors $\boldsymbol{f}$ and $\boldsymbol{g}$) , and that $\epsilon, \xi, \varsigma$, and $\mu $ are dimensionless. In the absence of an external excitation ($\boldsymbol{J}=0$) and assumption of non-dispersive (instantaneous) materials, Maxwell's equations become%
\begin{equation}
i\frac{\partial \boldsymbol{f}}{\partial t}=\hat{H}_{cl}\cdot \boldsymbol{f}  \label{ME1}
\end{equation}%
where $\hat{H}_{cl}=\boldsymbol{M}^{-1}\cdot \hat{N}$, which has the same form as the Schr\"{o}dinger equation with $\hbar =1$, where the operator $H_{cl}$ plays the role of a classical Hamiltonian. Because of this similarity between Maxwell's equations and the Schr\"{o}dinger equation it is straightforward to extend the Berry potential concept to electromagnetic
energy (photons); rather then, say, electrons acquiring a Berry phase while
transversing a path in parameters space, photons will do the same (the
polarization of the photon plays the role of particle spin). In this case,
we define $\boldsymbol{f}_{n}$ as a six-component eigenmode satisfying

\begin{equation}
\hat{H}_{cl}\cdot \boldsymbol{f}_{n}=E_{n}\boldsymbol{f}_{n}  \label{PEE1}
\end{equation}%
where $E_{n}=\omega _{n}$. Equation (\ref{PEE1}) applies also to lossless dispersive materials, which we consider in the following. Assuming the normalization condition $\left\langle
f_{n}|f_{m}\right\rangle =\delta _{nm}$, the Berry vector potential is

\begin{equation}
\mathbf{A}_{n}=i\left\langle \boldsymbol{f}_{n}|\nabla _{\mathbf{k}
}\boldsymbol{f}_{n}\right\rangle .  \label{PEE}
\end{equation}
such that $ \nabla_{\boldsymbol{\mathrm{k}}} $ operates over parameter space $ \boldsymbol{\mathrm{k}} = (k_x,~k_y,~k_z) $. In \cite{Haldane}, it was demonstrated by Raghu and Haldane that this result can be extended even for dispersive materials, $ \boldsymbol{M}{(\omega)} $, and the Berry vector potential takes the following form 

\begin{equation}\label{Eq:39}
\boldsymbol{A}_{nk} = Re \left \{ \frac{1}{2}i \boldsymbol{f} _{nk}^* \cdot \left[\frac{\partial (\omega \boldsymbol{M})}{\partial \omega}\right]_{\omega_{n,k}} \cdot \partial_k \boldsymbol{f}_{nk} \right \}
\end{equation}
with the inner product 

\begin{equation}
\left\langle \boldsymbol{f}_{n}|\boldsymbol{f}_{m}\right\rangle =\frac{1}{2}\boldsymbol{f}_{n}^{\ast }\frac{%
	\partial \left( \omega \boldsymbol{M}\left( \omega \right) \right) }{\partial \omega }%
\boldsymbol{f}_{m}
\end{equation}

We consider a magnetized plasma in the Voigt configuration (propagation perpendicular to the bias magnetic field $\mathbf{B}$), as depicted in Fig. 1 in the main text.

\section{Magnetically biased plasma continuum model}

For a single-component plasma biased with a static magnetic field $\mathbf{B}%
=\mathbf{z}B_{z}$, the permeability is $\mu=\mu_0$ and the relative permittivity has the form of a Hermitian antisymmetric tensor,
\begin{equation}\label{moe}
		\overline{\epsilon }=\left( 
		\begin{array}{ccc}
			\epsilon _{11} & \epsilon _{12} & 0 \\ 
			\epsilon _{21} & \epsilon _{22} & 0 \\ 
			0 & 0 & \epsilon _{33}%
		\end{array}%
		\right) 
\end{equation}%
where
\begin{align}
		\varepsilon _{11}& =\varepsilon _{22}=1-\frac{\omega _{p}^{2}}{\omega
			^{2}-\omega _{c}^{2}}\text{,  \: }\varepsilon _{33}=1-\frac{\omega _{p}^{2}}{%
			\omega ^{2}},\ \   \nonumber \\
		\varepsilon _{12}& =-\varepsilon _{21}=i\frac{-\omega _{c}\omega _{p}^{2}}{%
			\omega \left( \omega ^{2}-\omega _{c}^{2}\right) } \label{BMPM}
\end{align}%
where the cyclotron frequency is $\omega _{c}=\left( q_{e}/m_{e}\right)
B_{z}\ $and the plasma frequency is $\omega
_{p}^{2}=N_{e}q_{e}^{2}/\varepsilon _{0}m_{e}$. In the above, $N_{e}$ is the
free electron density, and $q_{e}$ and $m_{e}$ are the electron charge and
mass, respectivly.

The associated electromagnetic waves envelopes can be obtained by finding the solution $f=\left[\mathbf{E},\mathbf{H}\right] ^{T}$, of (\ref{PEE1}), $N \cdot f=\omega M \cdot f$, which is%
\begin{equation}
\left( 
\begin{array}{cc}
0 & -\mathbf{k}\times \mathbf{I}_{3\times 3} \\ 
\mathbf{k}\times \mathbf{I}_{3\times 3} & 0%
\end{array}%
\right) \cdot \left( 
\begin{array}{c}
\mathbf{E} \\ 
\mathbf{H}%
\end{array}%
\right) =\left( 
\begin{array}{cc}
\omega \epsilon _{0}\overline{\epsilon } & 0 \\ 
0 & \omega \mu _{0}\mathbf{I}_{3\times 3}%
\end{array}%
\right) \cdot \left( 
\begin{array}{c}
\mathbf{E} \\ 
\mathbf{H}%
\end{array}%
\right)   \label{EE4}
\end{equation}
so that 
\begin{equation}
\left( 
\begin{array}{cc}
-\mathbf{I}_{3\times 3} & -\frac{\overline{\epsilon }^{-1}}{\omega
	\epsilon _{0}}\cdot \mathbf{k}\times \mathbf{I}_{3\times 3} \\ 
\frac{1}{\omega \mu _{0}}\cdot \mathbf{k}\times \mathbf{I}_{3\times 3} & 
-\mathbf{I}_{3\times 3}%
\end{array}%
\right) \cdot \left( 
\begin{array}{c}
\mathbf{E} \\ 
\mathbf{H}%
\end{array}%
\right) =0.
\end{equation}%
With $\mathbf{H}=\widehat{\mathbf{z}}\rightarrow \mathbf{E}=\overline{%
	\mathbf{\epsilon }}^{-1}\cdot \frac{\widehat{\mathbf{z}}\times \mathbf{k}}{%
	\omega \epsilon _{0}}~~(\mathrm{TM}),~~\mathbf{E}=\widehat{\mathbf{z}}%
\rightarrow \mathbf{H}=\frac{\mathbf{k}}{\omega \mu _{0}}\times \widehat{%
	\mathbf{z}}~~(\mathrm{TE})$, we have the $6\times 1$ vectors%
\begin{align}
& \boldsymbol{f}_{nk}^{\text{TM}}=\left( 
\begin{array}{c}
\overline{\mathbf{\epsilon }}^{-1}\cdot \widehat{\mathbf{z}}\times \frac{%
	\mathbf{k}}{\epsilon _{0}\omega _{nk}} \\ 
\widehat{\mathbf{z}}%
\end{array}%
\right) ,  \nonumber \\
& \boldsymbol{f}_{nk}^{\text{TE}}=\left( 
\begin{array}{c}
\widehat{\mathbf{z}} \\ 
\frac{\mathbf{k}}{\mu _{0}\omega _{nk}}\times \widehat{\mathbf{z}}%
\end{array}%
\right) .~
\end{align}%
Because the envelopes of the electromagnetic waves in the above equations are
not normalized, the Berry potential is computed using

\begin{equation}
\mathbf{A}_{nk}=\frac{\text{Re}\{i\boldsymbol{f}_{nk}^{\ast }\cdot \frac{\partial }{%
		\partial \omega }(\omega \boldsymbol{M}{(\omega )}) \cdot \partial _{k}\boldsymbol{f}_{n,k}\}}{\boldsymbol{f}_{nk}^{\ast
	}\cdot \frac{\partial }{\partial \omega }(\omega \boldsymbol{M}{(\omega )}) \cdot \boldsymbol{f}_{n,k}}.
\label{Eq:60}
\end{equation}

Considering the Riemann sphere mapping of the $k_{x}-k_{y}$ plane as
detailed in \cite{Mario2}, it is possible to write the Chern number
associated with $n$th eigenmode branch as%
\begin{equation}
C_{n}=\frac{1}{2\pi }\int \mathbf{A}_{n,k=\infty }\cdot d\mathbf{l}-\frac{1}{2\pi 
}\int \mathbf{A}_{n,k=0^{+}}\cdot d\mathbf{l}
\end{equation}%
where the two line integrals are over infinite and infinitesimal radii
(north and south poles of the Riemann sphere), respectively. If we define $%
A_{nk}=\mathbf{A}_{nk}\cdot \hat{\mathbf{\phi}}$ then we have 
\begin{equation}
C_{n}=\lim\limits_{k\rightarrow \infty }(A_{n,\phi
	=0}k)-\lim\limits_{k\rightarrow 0^{+}}(A_{n,\phi =0}k).  \label{Eq:7}
\end{equation}

For a lossless TM-mode in propagating in the $x-y$ plane we have $k=k_{x}%
\hat{\mathbf{x}}+k_{y}\hat{\mathbf{y}}=k \cos(\phi )\hat{\mathbf{x}}+k \sin(\phi )\hat{\mathbf{y}}$. Writing%
\begin{equation}
\overline{\epsilon }^{-1}=\left( 
\begin{array}{ccc}
\alpha _{11} & \alpha _{12} & 0 \\ 
\alpha _{21} & \alpha _{22} & 0 \\ 
0 & 0 & \alpha _{33}%
\end{array}%
\right) 
\end{equation}%
we have%
\begin{equation}
f_{nk}=\left( 
\begin{array}{c}
\overline{\mathbf{\epsilon }}^{-1}\cdot \widehat{\mathbf{z}}\times \frac{%
	\mathbf{k}}{\epsilon _{0}\omega _{nk}} \\ 
\widehat{\mathbf{z}}%
\end{array}%
\right) =\left( 
\begin{array}{c}
\frac{-\alpha _{11}k_{y}+\alpha _{12}k_{x}}{\epsilon _{0}\omega _{n}} \\ 
\frac{-\alpha _{21}k_{y}+\alpha _{22}k_{x}}{\epsilon _{0}\omega _{n}} \\ 
0 \\ 
0 \\ 
0 \\ 
1%
\end{array}%
\right) ,~~\partial _{k}f_{nk}=\left( 
\begin{array}{c}
\frac{-\alpha _{11}\hat{y}+\alpha _{12}\hat{x}}{\epsilon _{0}\omega _{n}} \\ 
\frac{-\alpha _{21}\hat{y}+\alpha _{22}\hat{x}}{\epsilon _{0}\omega _{n}} \\ 
0 \\ 
0 \\ 
0 \\ 
0%
\end{array}%
\right) 
\end{equation}%
where%
\begin{equation}
\alpha _{11}=\frac{\epsilon _{22}}{\epsilon _{11}\epsilon _{22}-\epsilon
	_{12}\epsilon _{21}},~~\alpha _{22}=\frac{\epsilon _{11}}{\epsilon
	_{11}\epsilon _{22}-\epsilon _{12}\epsilon _{21}},~~\alpha _{12}=\frac{%
	-\epsilon _{12}}{\epsilon _{11}\epsilon _{22}-\epsilon _{12}\epsilon _{21}}%
,~~\alpha _{21}=\frac{-\epsilon _{21}}{\epsilon _{11}\epsilon _{22}-\epsilon
	_{12}\epsilon _{21}},
\end{equation}%
such that%
\begin{equation}
f_{nk}^{\ast }=\frac{1}{\epsilon _{0}\omega _{n}}\left( 
\begin{array}{cccccc}
(-\alpha _{11}k_{y}+\alpha _{12}k_{x})^{\ast } & (-\alpha _{21}k_{y}+\alpha
_{22}k_{x})^{\ast } & 0 & 0 & 0 & 1%
\end{array}%
\right) .
\end{equation}%
From the frequency derivative of the material response matrix, $\partial
_{\omega }(\omega M)$, we have $\beta _{ij}=\partial _{\omega }(\omega
\epsilon _{0}\epsilon _{ij})$. So, for the Berry potential we have%
\begin{equation}
\mathbf{A}_{nk}=\frac{\text{Re}\{i\boldsymbol{f}_{nk}^{\ast }\cdot \frac{\partial }{%
		\partial \omega }(\omega \boldsymbol{M}{(\omega )}) \cdot \partial _{k}\boldsymbol{f}_{n,k}\}}{\boldsymbol{f}_{nk}^{\ast
	}\cdot \frac{\partial }{\partial \omega }(\omega \boldsymbol{M}{(\omega )}) \cdot \boldsymbol{f}_{n,k}}=\frac{%
	Re\{N_{x}+N_{y}\}}{D}  \label{AB1}
\end{equation}%
where%
\begin{align}
& N_{x}=\frac{i}{2(\epsilon _{0}\omega _{n})^{2}}\{-2\alpha _{11}\alpha
_{12}[k_{x}\beta _{12}+k_{y}\beta _{11}]+(|\alpha _{11}|^{2}+|\alpha
_{12}|^{2})[k_{x}\beta _{11}-k_{y}\beta _{12}]\}\hat{x}  \label{AB2} \\
& N_{y}=\frac{i}{2(\epsilon _{0}\omega _{n})^{2}}\{2\alpha _{11}\alpha
_{12}[k_{x}\beta _{11}-k_{y}\beta _{12}]+(|\alpha _{11}|^{2}+|\alpha
_{12}|^{2})[k_{x}\beta _{12}+k_{y}\beta _{11}]\}\hat{y}  \nonumber \\
& D=\frac{|k|^{2}}{2(\epsilon _{0}\omega _{n})^{2}}[(|\alpha
_{11}|^{2}+|\alpha _{12}|^{2})\beta _{11}-2\alpha _{11}\alpha _{12}\beta
_{12}]+\mu_0 .
\end{align}%
Therefore, for the Chern number calculation we obtain%
\begin{align}
& A_{n}=\mathbf{A}_{n}\cdot \hat{\phi}=\frac{Re\{N_{y}cos(\phi
	)-N_{x}sin(\phi )\}}{D}  \label{BCa} \\
& A_{n}(\phi =0)=\frac{Re\{N_{y}(\phi =0)\}}{D},~~N_{y}(\phi =0)=\frac{ik}{%
	(\epsilon _{0}\omega _{n})^{2}}\{2\alpha _{11}\alpha _{12}\beta
_{11}+(|\alpha _{11}|^{2}+|\alpha _{12}|^{2})\beta _{12}\}  \nonumber \\
& A_{n}(\phi =0)k=\frac{Re(\frac{i|k|^{2}}{(\epsilon _{0}\omega _{n})^{2}}%
	\{2\alpha _{11}\alpha _{12}\beta _{11}+(|\alpha _{11}|^{2}+|\alpha
	_{12}|^{2})\beta _{12}\})}{\frac{|k|^{2}}{(\epsilon _{0}\omega _{n})^{2}}%
	[(|\alpha _{11}|^{2}+|\alpha _{12}|^{2})\beta _{11}-2\alpha _{11}\alpha
	_{12}\beta _{12}]+\mu_0 } \label{Eq:84}.
\end{align}%
These expressions are used below in calculating the Chern number from (\ref%
{Eq:7}).

Regarding to these parameters defined in previous section, the dispersion of the the TM mode is as follow

\begin{align}
&k^2 = \frac{\epsilon_{11}^2 + \epsilon_{12}^2 }{\epsilon_{11}} (\frac{\omega}{c})^2 \notag \\ &
k^2= \frac{\omega^2 (\omega^2 - \omega_c^2) -2 \omega^2 \omega_p^2 + \omega_p^4}{\omega^2 - \omega_c^2 - \omega_p^2} \frac{1}{c^2}
\end{align}

For the TM mode band as $ k= \frac{\omega^2 (\omega^2 - \omega_c^2) -2 \omega^2 \omega_p^2 + \omega_p^4}{\omega^2 - \omega_c^2 - \omega_p^2} \frac{1}{c^2} \rightarrow \infty $ we get the eigen frequency $ \omega_n \rightarrow \infty $ or the following eigen frequency

\begin{equation}\label{Eq:lowTM}
\omega^2 - \omega_c^2 - \omega_p^2 = 0 \rightarrow \omega_n=\sqrt{\omega_c^2 + \omega_p^2}
\end{equation}
such that $ \omega_n \rightarrow \infty $ belongs to the high frequency band and $ \omega_n=\sqrt{\omega_c^2 + \omega_p^2} $ belongs to the low frequency band.

For the TM mode if $ k= \frac{\omega^2 (\omega^2 - \omega_c^2) -2 \omega^2 \omega_p^2 + \omega_p^4}{\omega^2 - \omega_c^2 - \omega_p^2} \frac{1}{c^2} \rightarrow 0 $ we get the following eigen frequencies

\begin{align}\label{Eq:roots}
\omega^2 (\omega^2 - \omega_c^2) -2 \omega^2 \omega_p^2 + \omega_p^4 = 0~ \rightarrow ~ \begin{cases}
\omega_n^2= \frac{\omega_h^2}{2} \left \{ 1+ \sqrt{1-4(\frac{\omega_p}{\omega_h})^4}  \right \},& \text{for high frequency TM } \\
\omega_n^2= \frac{\omega_h^2}{2} \left \{ 1- \sqrt{1-4(\frac{\omega_p}{\omega_h})^4}  \right \},              & \text{for low frequency TM}
\end{cases}
\end{align}
such that $ \omega_h^2 = \omega_c^2 + 2\omega_p^2 $. It does worth to mention here that the roots in Eq. \ref{Eq:roots} which define the low and high frequency of the band gap are in fact the poles of $ \alpha_{11} $ and $\alpha_{12}$.

\subsection{Low frequency TM-band}

For the low frequency TM band when $ k \rightarrow \infty $ we have $ \omega_n = \sqrt{\omega_c^2 + \omega_p^2} $, $\epsilon_{11} = 0$ and $\alpha_{11} =0$. Therefore we get 

\begin{equation}
\lim_{k \rightarrow \infty}(A_{n,\phi=0}k) = Re \left \{ \frac{i \beta_{12}}{\beta_{11}} \right \}_{\omega_n=\sqrt{\omega_c^2 + \omega_p^2}} = -\frac{sgn(\omega_c)}{\sqrt{1+(\frac{\omega_p}{\omega_c})^2}}
\end{equation}

For the case of $k \rightarrow 0$, we have $ \omega_n^2= \frac{\omega_h^2}{2} \left \{ 1- \sqrt{1-4(\frac{\omega_p}{\omega_h})^4}  \right \}  $ which is the pole of $\alpha_{11}$ and $\alpha_{12}$ so $ \alpha_{11} \rightarrow \infty $, $ \alpha_{12} \rightarrow \infty  $, then for $\lim_{k \rightarrow 0}(A_{n,\phi=0}k)$ we get

\begin{align}
&\lim_{k \rightarrow 0}(A_{n,\phi=0}k) =  \lim_{k \rightarrow 0} \frac{ Re( \frac{i}{(\epsilon_0 c)^2} \{2 \frac{\alpha_{12}}{\alpha_{11}} \beta_{11} + (1 + \frac{|\alpha_{12}|^2}{\alpha_{11}^2}) \beta_{12} \} )   }{\frac{1}{(\epsilon_0 c)^2} \{ (1 + \frac{|\alpha_{12}|^2}{\alpha_{11}^2})\beta_{11} -2 \frac{\alpha_{12}}{\alpha_{11}}  \beta_{12} \}+ \frac{\mu}{\alpha_{11}}     }
\end{align}
in the above equation as $ \alpha_{11} \rightarrow \infty, ~ \alpha_{12} \rightarrow \infty $ then $ \frac{\alpha_{12}}{\alpha_{11}} = \frac{i\omega_c \omega_p^2/\omega}{\omega^2 - \omega_c^2 - \omega_p^2} $ and $ \frac{|\alpha_{12}|^2}{\alpha_{11}^2} = \frac{i\omega_c^2 \omega_p^4/\omega^2}{(\omega^2 - \omega_c^2 - \omega_p^2)^2} $ and $ \frac{\mu}{\alpha_{11}} \rightarrow 0 $ .  Therefore we have 

\begin{align}
&\lim_{k \rightarrow 0}(A_{n,\phi=0}k) = \left \{  \frac{ Re( \frac{i}{(\epsilon_0 c)^2} \{2 \frac{\alpha_{12}}{\alpha_{11}} \beta_{11} + (1 + \frac{|\alpha_{12}|^2}{\alpha_{11}^2}) \beta_{12} \} )   }{\frac{1}{(\epsilon_0 c)^2} \{ (1 + \frac{|\alpha_{12}|^2}{\alpha_{11}^2})\beta_{11} -2 \frac{\alpha_{12}}{\alpha_{11}}  \beta_{12} \}   } \right \} _{\omega_n^2= \frac{\omega_h^2}{2} \left \{ 1- \sqrt{1-4(\frac{\omega_p}{\omega_h})^4}  \right \}} = 1
\end{align}

Therefore the Chern number of the low frequency band is 

\begin{align}\label{Eq:Ch_H}
&C_n =  -\frac{sgn(\omega_c)}{\sqrt{1+(\frac{\omega_p}{\omega_c})^2}} -1
\end{align}

\subsection{High frequency TM-band}

For the high frequency band when $ k \rightarrow \infty $ we get $ \omega_n \rightarrow \infty $, $ \epsilon_{11} =1 $, $\epsilon_{12} = 0$, $ \alpha_{11} = 1 $, $ \alpha_{12}=0 $ and $ \beta_{12} = 0 $, so $\lim_{k \rightarrow \infty}(A_{n,\phi=0}k) = 0$ and for the case of $ k \rightarrow 0 $ we have $   \omega_n^2= \frac{\omega_h^2}{2} \left \{ 1+ \sqrt{1-4(\frac{\omega_p}{\omega_h})^4}  \right \}   $ again this is a pole of $ \alpha_{11} $ and $ \alpha_{12} $ so $ \alpha_{11} \rightarrow \infty, ~ \alpha_{12} \rightarrow \infty $ and we have 

\begin{align}
&\lim_{k \rightarrow 0}(A_{n,\phi=0}k) = \left \{  \frac{ Re( \frac{i}{(\epsilon_0 c)^2} \{2 \frac{\alpha_{12}}{\alpha_{11}} \beta_{11} + (1 + \frac{|\alpha_{12}|^2}{\alpha_{11}^2}) \beta_{12} \} )   }{\frac{1}{(\epsilon_0 c)^2} \{ (1 + \frac{|\alpha_{12}|^2}{\alpha_{11}^2})\beta_{11} -2 \frac{\alpha_{12}}{\alpha_{11}}  \beta_{12} \}   } \right \} _{\omega_n^2= \frac{\omega_h^2}{2} \left \{ 1+ \sqrt{1-4(\frac{\omega_p}{\omega_h})^4}  \right \}} = -1
\end{align}

Finally for the high frequency Chern number we have 

\begin{align}
&C_n =  0-(-1)=1
\end{align}

As it can be seen from Eq. \ref{Eq:Ch_H}, generally the Chern number of the low frequency TM band is non-integer but that of high frequency band is integer. 

\subsection{TE-Mode}

Using same procedure (the details of computation are omitted for conciseness) it is straightforward to show that for TE-Mode we have  

\begin{equation}\label{Eq:37}
\Delta C_n = \lim\limits_{k\rightarrow \infty} (A_{n,\phi=0}k) - \lim\limits_{k\rightarrow 0^+} (A_{n,\phi=0}k) = 0 
\end{equation}

\subsection{Wave vector cutoff for magneto-optic material response }

The emergence of a wave vector cutoff is well understood in
some materials. For example, a lossless electron gas described
by a drift-diffusion model has a spatially dispersive response
such that the permittivity seen by the transverse waves is $ \epsilon_T/\epsilon_0 = 1 - \omega_p^2 / \omega^2 $, whereas the permittivity seen by the
longitudinal waves is $ \epsilon_L / \epsilon_0 = 1 - \omega_p^2 / (\omega^2 - \nu^2 k^2) $, where $ \omega_p $ is the plasma frequency and $ \nu $ is a parameter with unities of velocity that depends on the diffusion coefficient \cite{Hanson1}. Hence, in the limit $ k \rightarrow \infty $ the longitudinal permittivity
approaches the response of the vacuum, i.e., the
response to longitudinal waves has a wave vector cutoff. In
this specific physical system, the wave vector cut-off for
longitudinal oscillations is a consequence of the diffusion
effects which act to avoid the localization of the electrons over
distances smaller than some characteristic diffusion length.
Inspired by this result, we may introduce a high-frequency
spatial cutoff by transforming a local material response as

\begin{equation}
\boldsymbol{M}_{reg}(\omega, \boldsymbol{k} ) = \boldsymbol{M}_{\infty} + \frac{1}{1+k^2/k_{max}^2} \left \{ \boldsymbol{M}(\omega) - \boldsymbol{M}_{\infty} \right\}   
\end{equation}
where $ \boldsymbol{M}_{\infty} = \lim\limits_{\omega \rightarrow \infty} \boldsymbol{M}(\omega) $. Based on this type of material response, the permittivity tensor components become

\begin{equation}
\epsilon_{11}(k)= \epsilon_{22}(k) = 1- \gamma \frac{\omega_p^2}{\omega^2 - \omega_c^2},~~ \epsilon_{12}(k) = - \epsilon_{21}(k) = -i \gamma \frac{\omega_c \omega_p^2}{ \omega (\omega^2 - \omega_c^2) } 
\end{equation}
such that $ \gamma = \frac{1}{1+k^2/k_{max}^2} $.

For this case, in dispersion equation $ k^2= k^2 = \frac{\epsilon_{11}(k)^2 + \epsilon_{12}(k)^2 }{\epsilon_{11}(k)} (\frac{\omega}{c})^2 $, we have $ k \rightarrow \infty $ if $ \epsilon_{11}(k)=0 $ or $ \omega_n \rightarrow \infty $, So the eigen frequency of the higher TM band is $ \omega_n \rightarrow \infty $ and that of lower frequency band comes from the zero of $ \epsilon_{11}(k) $.

\begin{align}
\epsilon_{11}(k) = 1- \gamma \frac{\omega_p^2}{\omega^2 - \omega_c^2} = 0 \rightarrow \omega_n = \sqrt{\omega_c^2 + \gamma \omega_p^2} 
\end{align}
when $ k \rightarrow \infty $ then $ \gamma \rightarrow 0 $ so for low frequency band eigen frequency we get $ \omega_n = \lim\limits_{\gamma \rightarrow 0} \sqrt{\omega_c^2 + \gamma \omega_p^2}  = |\omega_c| $.

For the case of $ k \rightarrow 0 $ ( $ \gamma \rightarrow 1 $) we have same dispersion equation when we had no wave vector cut-off and the eigen frequencies are going to be those in Eq. 4. Therefore nothing changes in this case.

As it was mentioned, For low frequency TM band as $ k \rightarrow \infty ~ (\gamma \rightarrow 0) $ we have $ \omega_n = |\omega_c| $ so $ \epsilon_{11}(k) = 0 $, $ \alpha_{11}(k) = 0 $ and we get

\begin{equation}
\lim_{k \rightarrow \infty}(A_{n,\phi=0}k) = Re \left \{ \frac{i \beta_{12}(k)}{\beta_{11}(k)} \right \}_{\omega_n=|\omega_c|}
\end{equation}
such that $ \beta_{11}(k) = 1+\gamma \omega_p^2 \frac{\omega^2 + \omega_c^2}{(\omega^2 - \omega_c^2)^2}, ~~ \beta_{12} = 2i \gamma \omega_c \omega_p^2 \frac{\omega}{(\omega^2 - \omega_c^2)^2}  $ so the contribution form $ k \rightarrow \infty $ in low frequency TM band is 

\begin{equation}
\lim_{k \rightarrow \infty,~\gamma \rightarrow 0}(A_{n,\phi=0}k) = Re \left \{ \frac{i \beta_{12}(k)}{\beta_{11}(k)} \right \}_{\omega_n=|\omega_c|} = \lim\limits_{\gamma \rightarrow 0} \frac{-2 \gamma \omega_c^2 \omega_p^2}{1+2 \gamma \omega_c^2 \omega_p^2 } = 0
\end{equation}

For this band if $ k \rightarrow 0~ (\gamma \rightarrow 1) $ then wave vector cut-off does nothing and we have 

\begin{align}
&\lim_{k \rightarrow 0}(A_{n,\phi=0}k) = \left \{  \frac{ Re( \frac{i}{(\epsilon_0 c)^2} \{2 \frac{\alpha_{12}}{\alpha_{11}} \beta_{11} + (1 + \frac{|\alpha_{12}|^2}{\alpha_{11}^2}) \beta_{12} \} )   }{\frac{1}{(\epsilon_0 c)^2} \{ (1 + \frac{|\alpha_{12}|^2}{\alpha_{11}^2})\beta_{11} -2 \frac{\alpha_{12}}{\alpha_{11}}  \beta_{12} \}   } \right \} _{\omega_n^2= \frac{\omega_h^2}{2} \left \{ 1- \sqrt{1-4(\frac{\omega_p}{\omega_h})^4}  \right \}} = 1
\end{align}
so for low frequency band we get following Chern number

\begin{align}
&C_n =  0-1=-1
\end{align}

For the high frequency TM band as $ k \rightarrow \infty~ (\gamma \rightarrow 0) $, as it was explained before, we get $ \lim_{k \rightarrow \infty}(A_{n,\phi=0}k) = 0 $ and when $ k \rightarrow 0~ (\gamma \rightarrow 1) $ again we have

\begin{align}
&\lim_{k \rightarrow 0}(A_{n,\phi=0}k) = \left \{  \frac{ Re( \frac{i}{(\epsilon_0 c)^2} \{2 \frac{\alpha_{12}}{\alpha_{11}} \beta_{11} + (1 + \frac{|\alpha_{12}|^2}{\alpha_{11}^2}) \beta_{12} \} )   }{\frac{1}{(\epsilon_0 c)^2} \{ (1 + \frac{|\alpha_{12}|^2}{\alpha_{11}^2})\beta_{11} -2 \frac{\alpha_{12}}{\alpha_{11}}  \beta_{12} \}   } \right \} _{\omega_n^2= \frac{\omega_h^2}{2} \left \{ 1+ \sqrt{1-4(\frac{\omega_p}{\omega_h})^4}  \right \}} = -1
\end{align}

so for high frequency band we get following Chern number

\begin{align}
&C_n =  0-(-1)=+1
\end{align}